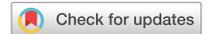

OPEN

# A survey on computational spectral reconstruction methods from RGB to hyperspectral imaging

Jingang Zhang[1,7], Runmu Su[1,2,7], Qiang Fu[3], Wenqi Ren[4], Felix Heide[5] & Yunfeng Nie[6]✉

Hyperspectral imaging enables many versatile applications for its competence in capturing abundant spatial and spectral information, which is crucial for identifying substances. However, the devices for acquiring hyperspectral images are typically expensive and very complicated, hindering the promotion of their application in consumer electronics, such as daily food inspection and point-of-care medical screening, etc. Recently, many computational spectral imaging methods have been proposed by directly reconstructing the hyperspectral information from widely available RGB images. These reconstruction methods can exclude the usage of burdensome spectral camera hardware while keeping a high spectral resolution and imaging performance. We present a thorough investigation of more than 25 state-of-the-art spectral reconstruction methods which are categorized as prior-based and data-driven methods. Simulations on open-source datasets show that prior-based methods are more suitable for rare data situations, while data-driven methods can unleash the full potential of deep learning in big data cases. We have identified current challenges faced by those methods (e.g., loss function, spectral accuracy, data generalization) and summarized a few trends for future work. With the rapid expansion in datasets and the advent of more advanced neural networks, learnable methods with fine feature representation abilities are very promising. This comprehensive review can serve as a fruitful reference source for peer researchers, thus paving the way for the development of computational hyperspectral imaging.

Hyperspectral imaging refers to the dense sampling of spectral features with many narrow bands. Unlike traditional RGB images, each pixel of hyperspectral images (HSIs) contains a continuous spectral curve to identify the substance of the corresponding objects. Since spectral information is valuable to distinguish materials, hyperspectral imaging has been frequently used in scientific research, such as remote sensing[1,2], agriculture[3], geology[4,5], astronomy[6], and medical imaging[7,8], just to name a few. By virtue of the highly recognized advantages as a non-contact, non-destructive detection method, hyperspectral imaging has attracted considerable attention and intensive research. However, the devices for acquiring HSIs are typically more complicated and expensive than common cameras[9–11]. Figure 1 illustrates that many more optical components have to be employed in hyperspectral imagers than common cameras. In addition, most hyperspectral imagers rely on precise scanning (e.g., pushbroom or whiskbroom scanners) to generate 3D datacube (2D spatial and 1D spectral information), which hinders them from being portable and cost-effective. Although some strategies have been proposed recently to achieve snapshot hyperspectral imaging without scanning[12,13], the effective spatial/spectral resolution is low. From this perspective, it seems that reducing opto-mechanical components in hyperspectral systems is usually a disturbing compromise of system performance and cost.

In the past decade, the computer vision community has achieved tremendous success in semantic understanding of visual information with HSIs, such as image segmentation[14,15], recognition[16,17], tracking[18,19], pedestrian detection[20], and anomaly detection[21]. The lack of mainstream, easy-operational devices to capture HSIs quickly becomes a bottleneck in further research. In order to obtain spectral information more effectively, a recent trend is to extract spectral information from RGB images. Given the fact that consumer cameras have become worldwide prevailing in daily life, and a huge amount of high-quality color images can be captured easily and very cost-effective, the success of such methods could bring extensive benefits for HSIs to be applied in previously limited circumstances. For example, endoscopes[22] that work in a narrow space could harness the abundant existence of endoscopic RGB images to extend their applications in hyperspectral cancer detection[23]. Industrial machine vision cameras could be boosted for increased food quality inspection using spectral features[24]. Consumer-level point-of-care applications could be enabled by converting mobile phone cameras into medical apparatus for vein localization and visualization[25].

[1]School of Future Technology, University of Chinese Academy of Sciences, Beijing 100039, China. [2]School of Computer Science and Technology, Xidian University, Xi'an 710071, China. [3]King Abdullah University of Science and Technology, Thuwal 23955-6900, Saudi Arabia. [4]State Key Laboratory of Information Security, Institute of Information Engineering, Chinese Academy of Sciences, Beijing 100093, China. [5]Computational Imaging Lab, Princeton University, Princeton, NJ 08544, USA. [6]Department of Applied Physics and Photonics, Vrije Universiteit Brussel, 1050 Brussels, Belgium. [7]These authors contributed equally: Jingang Zhang and Runmu Su. ✉email: Yunfeng.Nie@vub.be

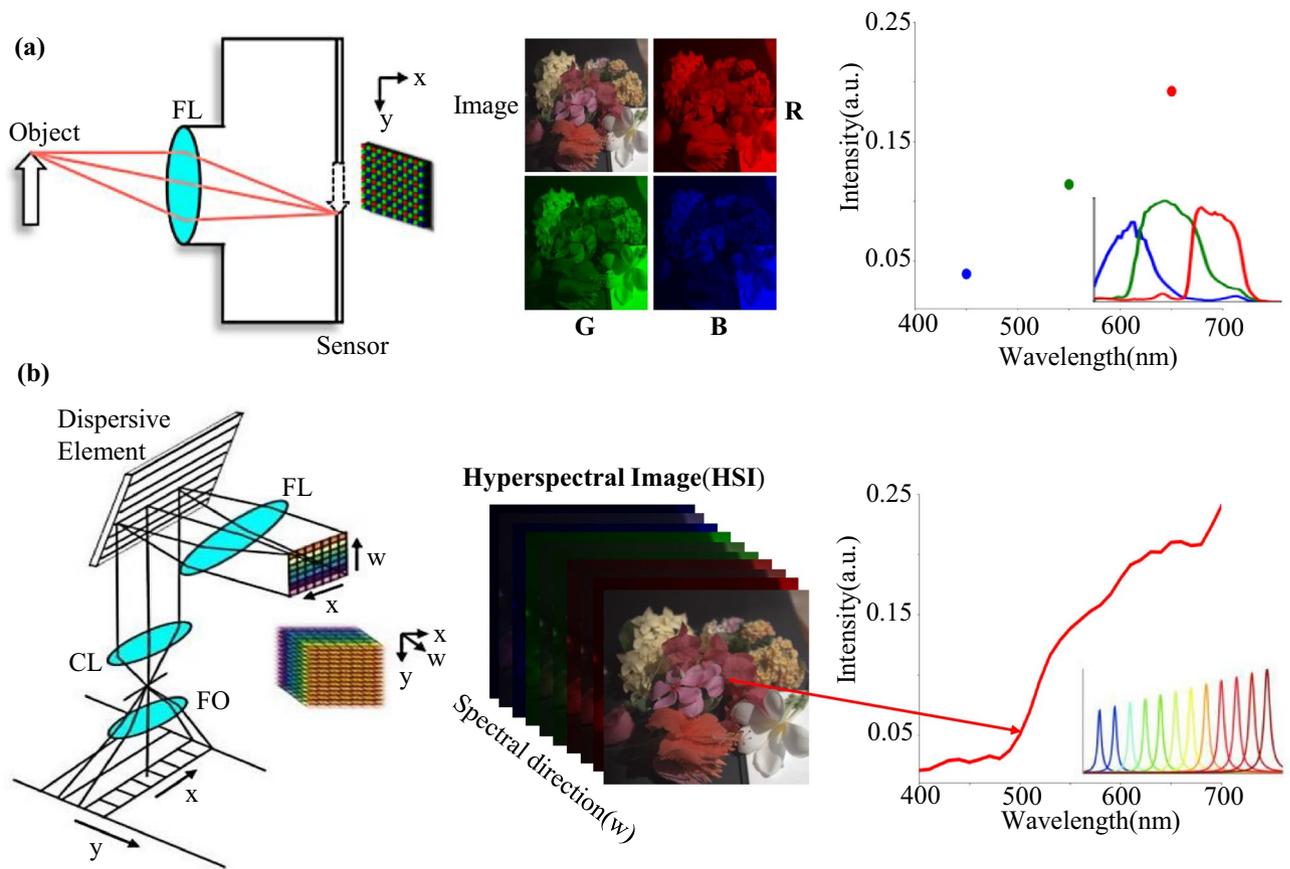

**Figure 1.** Schematic diagrams of an RGB camera (**a**) and a typical hyperspectral imager (**b**). In an RGB image, each pixel is combined with three discrete color values, which is integrated from wide R, G, B spectra. In a hyperspectral image, each pixel is a continuous spectral curve that is filtered from a series of narrow spectral bands. *FL* focusing lens group, *FO* front objective, *CL* collimating lens group. The image data are from the KAUST-HS open-source dataset[26].

However, recovering hyperspectral information from RGB images is an ill-posed inverse problem. In order to encourage tackling this problem, two competitions on spectral reconstruction from RGB images (NTIRE-2018[27] and NTIRE-2020[28]) have been launched, resulting in many brilliant methods to promote the most competitive solutions. Typically, they can be categorized into two types, prior-based and data-driven methods. Both methods take advantage of the redundancy in image data, whereas they differ from each other in the principles. Prior-based methods explore statistical information (priors) in HSIs, e.g., sparsity[29], spatial structure similarity and spectral correlation[30], to seek plausible solutions in a more constrained space. Data-driven methods exploit advantages of the abstract features in large-scale RGB and hyperspectral image datesets, rather than handmade priors, to find more accurate solutions. Various neural network architectures have been proposed to increase the reconstruction accuracy, such as convolutional neural network (CNN)[31] and Generative Adversarial Network (GAN)[32].

As the hyperspectral reconstruction from RGB images opens up more and more promising applications, a thorough review becomes a necessity to inspire future research. In this work, we take a comprehensive survey on state-of-the-art computational methods that reconstruct hyperspectral images from RGB images. We first introduce the fundamental image formation models for RGB and hyperspectral images. A list of HSI datasets and performance metrics used in those methods are then briefly described. A detailed review on the prior-based methods and data-driven methods are explained and compared. We also systematically compare several representative algorithms in simulation using available open-source datasets. Finally, the challenges and trends faced by current methods are summarized for inspiration of future work on this topic.

| Dataset | Amount | Resolution | Spectral channels | Spectrum/(nm) | Featured scenes |
|---|---|---|---|---|---|
| CAVE[34] | 32 | 512 × 512 | 31 | 400–700 | Skin, hair, food and drink |
| ICVL[29] | 203 | 1392 × 1300 | 31 | 400–700 | Urban, rural, indoor and plant |
| BGU-HS[27] | 286 | 1392 × 1300 | 31 | 400–700 | Urban, rural, indoor and plant |
| ARAD-HS[28] | 510 | 512 × 482 | 31 | 400–700 | Statue, vehicle and paint |
| KAUST-HS[26] | 409 | 512 × 512 | 34 | 400–730 | Vehicle, food, building and toy |

**Table 1.** Properties of five open-source hyperspectral datasets.

## Fundamentals

**Image formation model.** An RGB image is usually generated in three wide spectral channels with Bayer filters, while an HSI has dozens of narrow spectral bands, as shown in Fig. 1. In principle, an HSI is obtained by the interaction between the spectral reflectance and the illumination spectrum. The spectral reflectance is an essential physical attributes, thus the HSI can be used to identify substances. According to the Lambertian assumption[33], the relation between HSIs and RGB images is

$$I_c(x,y) = \int_{w_1}^{w_2} R(x,y,w)L(w)S_c(w)dw, \quad (1)$$

where $I_c$ represents each channel of an RGB image (c = R, G, B), $S_c(w)$ denotes the spectral response function (SRF), $R(x,y,w)$ is the spectral reflectance of the objects, and $L(w)$ is the illumination spectrum. A hyperspectral image $H(x,y,w)$ is typically defined as the multiplication of the scene's spectral reflectance and the illumination spectrum, so Eq. (1) becomes

$$I_c(x,y) = \int_{w_1}^{w_2} H(x,y,w)S_c(w)dw. \quad (2)$$

The challenge in spectral reconstruction (SR) is to invert the above forward image formation model in an optimization framework to find the multi-channel data-cube $H(x,y,w)$, where only the captured 3-channel RGB image $I_c$ is known. Further, when the illumination spectrum $L(w)$ is known or calibrated, the scene reflectance $R(x,y,w)$ can be calculated accordingly.

**Datasets and performance evaluation metrics.** Data and performance metrics have been playing more and more significant roles with the advent of artificial intelligence. A couple of emerging HSI datasets have been available in recent years for training and verifying deep learning networks. We first introduce available open-source datasets that help boosting the research, and then discuss various performance metrics for the evaluation and comparison of different algorithms.

*Open-source datasets.* Table 1 lists five HSI datasets commonly used in the SR community, all of which have a small or medium size. As this technique advances, more HSIs can be captured or collected, enabling larger datasets to be available. The most important attributes of the existing datasets are data amount, spatial resolution, spectral channels, and the diversity of the scenes. The detailed properties of these datasets are explained below.

- CAVE[34] is an early and frequently used hyperspectral dataset. Unlike others, this dataset was captured using a tunable filter instead of a dispersive grating or prism. It contains 32 images with 512 × 512 pixels and 31 spectral bands between 400 nm and 700 nm. CAVE is a collection of various indoor objects with controlled illumination.
- ICVL[29] is collected and published by Arad and Ben Shahar in 2016. This dataset contains 203 images acquired by a line-scanning camera (Specim PS Kappa DX4 hyper-spectrometer). Various indoor and outdoor scenes are included to increase the diversity. The spatial resolution is 1392 × 1300, and 31 spectral channels from 400 nm to 700 nm in 10 nm interval are publicized for visible applications.
- BGU-HS[27] has been the largest natural HSI dataset so far. During the SR challenge NTIRE-2018, the dataset has been expanded to include 286 images, further divided into 256 training images, 10 verification images, and 20 test images. Each HSI has a spatial resolution of 1392 × 1300, and 31 spectral bands, ranging from 400 to 700 nm with an interval of 10 nm.
- ARAD-HS[28] is the an HSI dataset for NTIRE-2020 with 510 images, further divided into 450, 30 and 30 images for training, validation, and testing respectively. The spatial resolution is 512 × 482, and the number of spectral band is 31. This dataset was collected by a portable hyperspectral camera (Specim-IQ). A large variety of indoor and outdoor scenes are collected, such as statues, vehicles and paints.
- KAUST-HS[26] is an HSI dataset with 409 spectral reflectance images used originally for illumination estimation. The spatial dimension is 512 × 512. Its spectral range covers from 400 to 730 nm, with a spectral interval of 10 nm. The dataset was captured with Specim-IQ. Diverse indoor scenes (clothes, toys, vegetables) and outdoor scenes (buildings, plants, vehicles) are included. This dataset differs from other datasets in that the reflectance HSIs are calibrated by a standard white board, such that the illumination spectrum $L(w)$ is removed.

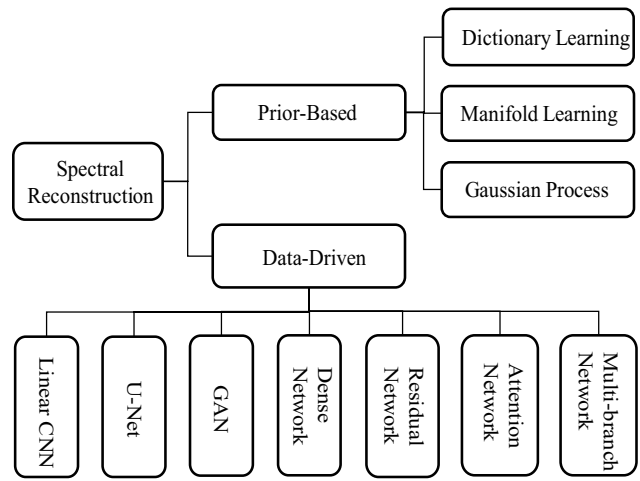

**Figure 2.** Classification of state-of-the-art spectral reconstruction methods.

*Performance evaluation metrics.* Spatial and spectral errors are important measures to evaluate the reconstructed HSIs compared with their ground truths. Three major metrics are commonly used in the literature, namely Mean Relative Absolute Error (MRAE)[28], Root Mean Square Error (RMSE)[28], and Spectral Angle Mapper (SAM)[35]. They are defined as follows,

$$\text{MRAE} = \frac{1}{N} \sum_{i=1}^{N} \left( \frac{|H_{GT}^i - H_{SR}^i|}{H_{GT}^i} \right), \tag{3}$$

$$\text{RMSE} = \sqrt{\frac{1}{N} \sum_{i=1}^{N} \left(H_{GT}^i - H_{SR}^i\right)^2}, \tag{4}$$

where $H_{GT}^i$ and $H_{SR}^i$ are the i-th pixel value of the ground truth and the reconstructed spectral image respectively, and $N$ is total number of pixels.

$$\text{SAM} = \frac{1}{M} \arccos \left( \sum_{j=1}^{M} \frac{\left(S_{GT}^j\right)^T \cdot S_{SR}^j}{\left\|S_{GT}^j\right\|_2 \cdot \left\|S_{SR}^j\right\|_2} \right), \tag{5}$$

where $S_{GT}^j$ and $S_{SR}^j$ are the 1D spectral curve of the j-th hyperspectral pixel in the ground-truth and reconstructed HSIs, and $M$ is the total number of pixels in an image slice. MRAE and RMSE indicate the numerical errors of the reconstructed images. SAM reflects the spectral similarity between the reconstructed and the original spectra at the pixel level. Although these metrics have been extensively adopted, they do suffer from some drawbacks. For example, a large spectral deviation for narrow-band objects could be easily averaged out if the rest of the wide-band scene is well reconstructed. More accurate and comprehensive HSI image evaluation metrics are worth studying further.

## Algorithm survey

Given the main principles, we divide the SR algorithms into two categories: prior-based and data-driven methods. Prior knowledge is the statistical information in the data to represent the inherent attributes and characteristics of the image. Among prior-based methods, they are classified into three types by the difference in their prior selections. We investigate various network models in the data-driven methods and categorize them into several groups, namely Linear CNN, U-Net model, GAN model, etc. The overall taxonomy of the algorithms to be analyzed are shown in Fig. 2.

**Prior-based methods.** The target is to reconstruct an HSI $\widetilde{\mathbf{H}} \in \mathbb{R}^{m \times n \times C}$ from an RGB image $\widetilde{\mathbf{Y}} \in \mathbb{R}^{m \times n \times c}$, where $m$ and $n$ are the number of pixels in the 2D spatial dimensions, $C$ is the number of spectral band and $c$ represents R, G and B channels. To simplify the process, we use a 2D matrix to represent the image, where $\mathbf{Y}$, $\mathbf{H}$ respectively are written as $\mathbf{Y} \in \mathbb{R}^{c \times N}$ and $\mathbf{H} \in \mathbb{R}^{C \times N}$, and $N = mn$ is the total number of pixels within an image slice. In this case, each column represents one spectral curve, and each row corresponds to the reshaped image of one certain spectral band.

| Category | Methods | Priors |
|---|---|---|
| Dictionary learning | Sparse coding[29] | Sparsity |
| | SR A+[37] | Sparsity, local euclidean linearity |
| | Multiple non-negative sparse dictionaries[36] | Spatial structure similarity, spectral correlation |
| | Local linear embedding sparse dictionary[38] | Color and texture, local linearity |
| | Spatially constrained dictionary learning[39] | Spatial context |
| Manifold learning | SR manifold mapping[40] | Low-dimensional manifold |
| Gaussian process | SR Gaussian process[30] | Spectral physics, spatial structure similarity |

**Table 2.** An overview of prior-based spectral reconstruction methods.

As we know that, each pixel in an HSI datacube represents a combination of different substances with certain proportions. Thus, the spectrum of each pixel is the linear combination of the basis spectra of those substances (also called endmember) and the corresponding proportions (called abundance), described as $\mathbf{H}_i = \mathbf{EA}_i$, where $\mathbf{E}$ refers to the basis spectra matrix, and $\mathbf{A}_i$ denotes the proportion vector on each pixel. The RGB image is obtained by applying the spectral response function (SRF) of the detector $\mathbf{S}$ to the HSI, which is described as $\mathbf{Y}_i = \mathbf{SH}_i$ or $\mathbf{Y}_i = \mathbf{SEA}_i$. From this equation, recovering the spectral information $\mathbf{E}$, $\mathbf{A}_i$ from the RGB image $\mathbf{Y}_i$ is an inverse problem, and the solution is not unique. In order to narrow down the solution space, we need to use prior knowledge as constraints. Therefore, computational reconstruction of HSIs from RGB images is essentially to solve the following optimization problem

$$\arg\min_{\mathbf{E},\mathbf{A}_i} \|\mathbf{H}_i - \mathbf{EA}_i\|^2 + \|\mathbf{Y}_i - \mathbf{SEA}_i\|^2 + \lambda P(\mathbf{A}_{i,j}), \quad i,j = 1,2\ldots,N \text{ and } i \neq j, \qquad (6)$$

where $P(\mathbf{A}_{i,j})$ is the prior term, and $\lambda$ denotes the weighting factor. Prior knowledge about HSI usually includes sparsity[29], spatial structure similarity[30], correlation between spectra[36], etc. Typical methods and their priors are summarized in Table 2.

*Dictionary learning.* Statistical analysis of HSIs shows that they are sparse in space and spectrum[41], and the spectral information is expressed as a sparse combination of basis spectra. The dictionary is introduced to store the basis spectra $\mathbf{E}$, and the coefficients (analogous to the proportions $\mathbf{A}$) is further obtained by dictionary learning. The core of most dictionary learning methods is to solve the basis spectra or the spectral dictionary. Once obtained, it is projected into the RGB space via the SRF to form an RGB dictionary. The last step is to recover the spectral information of an input RGB image using the RGB dictionary. These methods may not work well when the SRFs have to be estimated. For example, the real SRF might be different from the vendor's data, or it is absent when the cameras that take the RGB images are unknown. The finding of new prior knowledge, e.g., the spatial structure similarity and high correlation across spectra, is used as proper regularization, which leads to SR dictionary learning with enhanced performance.

Sparse coding. Sparse coding is a dictionary learning method that focuses on finding the basis spectra (spectral sub-dictionaries) and their coefficients to represent the latent HSI. As the basis spectra are overcomplete, the coefficients are guaranteed to be sparse. Based on this sparsity of HSI coefficients, Arad et al.[29] proposed this method to obtain the spectral dictionary, according to the inverse mapping problem in Eq. (6). This method is fast when the dataset is small. As the dataset increases, the dictionary capacity can be enlarged; as a result, the reconstruction time is also prolonged. Nevertheless, this method does not take into account the spatial correlation, so the quality of the reconstructed image is limited.

SR A+. The SR A+ method proposed by Aeschbacher et al.[42] is the same as the sparse coding method in obtaining the hyperspectral and RGB dictionaries. The difference is that SR A+ establishes the RGB-to-hyperspectral mapping in a local dictionary, but not a global one as in sparse coding. More specifically, it solves the mapping coefficients from neighboring anchor points by a least-square optimization, which runs faster with a higher accuracy. However, this method is also pixel-wise and does not consider the spatial structure similarity.

Multiple non-negative sparse dictionaries. Fu et al.[36] proposed this method to reconstruct the spectral reflectance and the illumination spectrum simultaneously. It clusters the spectral reflections in spectral datasets and independently learns a specific spectral dictionary that has multiple non-negative sparse feature. Unlike the above-mentioned methods, the multiple non-negative sparse prior is used to provide a more compact base representation for each cluster, effectively describing the spectral information of various substances in the scene. Consequently, this method can greatly improve the performance of sparse representation.

Local linear embedding sparse dictionary. The above sparse dictionary only considers the sparsity of spectral information, but not local linearity. This lowers the reconstruction accuracy, and the reconstructed image is affected by metamerism. Li et al.[38] proposed the Local Linear Embedding Sparse Dictionary method to solve this problem. In this method, the locally best samples are selected to reduce the redundancy of the samples in a global space. In the process of dictionary learning, the local linearity of the spectrum is introduced to make the dictionary compact and efficient. Besides, the texture information is also considered to ensure the reconstructed HSI quality and reduce metamer[43,44]. Although this method uses these features to improve the reconstruction performance, they are all hand-crafted assumptions.

*Spatially constrained dictionary learning.* In the early stage, the SR methods based on dictionary learning are pixel-wise, so the reconstruction results are not sufficiently accurate. Geng et al.[39] introduced spatial context information into sparse representation, which leads to a spatially constrained dictionary learning method. Compared with the pixel-wise reconstruction methods, the introduced spatial context information can preserve the spatial structures and the physical connections in the image.

*Manifold learning.* Manifold learning is to find a low-dimensional manifold[45–50] that can uniquely represent high-dimensional data. In this sense, hyperspectral data can be represented by a set of low-dimensional manifolds. With this prior knowledge, an SR model based on manifold learning is established.

*SR manifold mapping.* Jia et al. proposed a manifold learning method[40] to simplify the three-to-many mapping problem into a three-to-three problem. An isometric mapping algorithm[51] is used to reduce the dimensionality of the basis spectra to a three-dimensional subspace. This method can explore the systematic characteristics of hyperspectral images, which constrains the solution space to obtain more accurate solution. So far, the method is merely studied under the conditions of known illumination (sunlight) and known camera parameters.

*Gaussian process.* The base spectra are smooth functions of the wavelengths, so they can be approximately modelled by Gaussian functions or processes. Following this logic, the Gaussian processes[52] are used to reconstruct spectral information from RGB images.

*SR Gaussian process.* This method uses Gaussian process to model the basis spectra of different substances[30], and interpolates them with the RGB image and the detector SRF to obtain the corresponding coefficients. If applied to Eq. (6), it tries to solve a set of Gaussian processes **E** and the corresponding coefficients **A**. K-means clustering method has been used to find similar image patches, thus introducing the spatial similarity and spectral correlation to speed up the program. Meanwhile, the physical characteristics of the spectral signals are incorporated into the Gaussian processes through its kernel and the use of non-negative mean prior probability distribution, which improves the reconstruction accuracy. The drawback is that the model is complex with numerous parameters, and very difficult to apply in more general cases.

**Data-driven methods.** The prior-based methods are almost hand-crafted and rely heavily on the selected priors and known SRFs. In order to overcome these limitations, deep learning methods are proposed. From the perspective of the most distinctive features in the network architectures, we divide data-driven deep learning based methods into seven groups, as seen in Table 3.

*Linear CNN.* CNN has been applied a lot in the field of SR. Linear CNN is a stack of convolutional layers, and the input sequentially flows from the initial layer to the later layers. This network architecture only has a single path and does not include multiple branches. Note that some linear CNNs learn to reproduce residual images called low-spectral image (LSI) compared to HSI, which does not add any branch as in the residual network. Main linear CNN-based SR methods are described as follows.

HSCNN. HSCNN[31] is a unified deep learning framework for restoring hyperspectral information from spectrally undersampled images, such as RGB and compressed sensing[68,69] images. HSCNN inherits the spatial super-resolution algorithm VDSR[70]. The difference between HSCNN and VDSR is in the first and last layers. The other layers have the same configuration as the VDSR, as shown Fig. 3. Mean square error $\mathscr{L}_{MSE}$ is used to train HSCNN. Although HSCNN has a simple network structure, it can achieve good reconstruction fidelity. One limitation of HSCNN is that the SRF should be known in the spectral upsampling operation. Besides, HSCNN can fail to improve performances even if the network depth increases.

SR2D/3DNet. SR2D/3DNet[53] is a representative solution from the challenge of NTIRE-2018[27]. The authors use 2D- and 3D-convolution kernel based CNN to solve the SR problem. The difference is that 2DNet operates independently on different channels and only considers the spatial domain, while 3DNet considers the relations among multiple channels. The two network structure diagrams are shown in Fig. 3. The 2D/3D network is trained with mean absolute error $\mathscr{L}_{MAE}$ using the paired image patches. This method jointly learns the correlation between spatial and spectral domains, which slightly improves network performance. Due to a low network complexity, it is difficult to extract effective features.

Residual HSRCNN. This network was evolved from SRCNN[71] and HSRCNN[54]. HSRCNN has three convolutional layers that reconstruct low-frequency information of spectral images. Since it is difficult for HSRCNN to recover high-frequency or residual spectral information, the authors overwrite the baseline CNN of HSRCNN to form Residual HSRCNN, as shown in Fig. 3. The outputs of HSRCNN and Residual HSRCNN are added together to form the final reconstructed HSI. During the training, the RGB image is divided into 15 × 15 overlapping image patches as the network input, and the Euclidean loss $\mathscr{L}_{eu}$ is used. This method learns a residual mapping that generates the difference between LSIs and HSIs. It avoids learning a complicated transformation from RGB images to hyperspectral images, instead only requires learning a residual map to restore the missing high-frequency details. However, it is still challenging to restore highly accurate spectra.

| Category | Methods | Depth | Filters | Loss function | Framework | Optimizer |
|---|---|---|---|---|---|---|
| Linear CNN | HSCNN[31] | 5 | 64 | $\mathcal{L}_{MSE}$ | Caffe | Adam |
| | SR2D/3DNet[53] | 5 | 64 | $\mathcal{L}_{MAE}$ | Keras/Tensorflow | Adam |
| | Residual HSRCNN[54] | 6 | 64 | $\mathcal{L}_{eu}$ | Caffe | SGD |
| U-Net | SRUNet[55] | 5 | 128 | $L_{\Delta E}$ | Pytorch | SGD |
| | SRMSCNN[35] | 10 | 1024 | $\mathcal{L}_{MSE}$ | Pytorch | Adam |
| | SRMXRUNet[56] | 56 | 4096 | $\mathcal{L}_{mper}$ | FastAI | AdamW |
| | SRBFWU-Net[57] | 4 | 512 | $\mathcal{L}_{MRAE}, \mathcal{L}_1$ | Pytorch | Adam |
| GAN | SRCGAN[32] | 8 | 512 | $\mathcal{L}_{GAN}, \mathcal{L}_1$ | Keras | Adam |
| | SAGAN[58] | 10 | 1024 | $\mathcal{L}_{adv}, \mathcal{L}_1$ | Pytorch | Adam |
| Dense network | SRTiramisuNet[59] | 23 | 16 | $\mathcal{L}_{eu}$ | Keras | Adam |
| | HSCNN+[60] | 160 | 64 | $\mathcal{L}_{MRAE}$ | Pytorch/Tensorflow | Adam |
| Residual network | SREfficientNet[61] | 9 | 128 | $\mathcal{L}_2$ | Tensorflow | Adam |
| | SREfficientNet+[62] | 21 | 128 | $\mathcal{L}_2$ | Tensorflow | Adam |
| Attention network | SRAWAN[63] | 61 | 200 | $\mathcal{L}_{MRAE}, \mathcal{L}_{CSS}$ | Pytorch | Adam |
| | SRHRNet[64] | 57 | 256 | $\mathcal{L}_1$ | Pytorch | Adam |
| | SRRPAN[65] | 135 | 64 | $\mathcal{L}_{MRAE}$ | Pytorch | Adam |
| Multi-branch network | SRLWRDNet[66] | 40 | 32 | $\mathcal{L}_2, \mathcal{L}_{SSIM}$ | Keras | Adam |
| | SRPFMNet[67] | 9 | 64 | $\mathcal{L}_1$ | Pytorch | Adam |

**Table 3.** Overview of data-driven deep learning methods. Depth is the number of convolutional layers. Filters the number of convolution kernels. Framework is the original platform used in the references without excluding the usage of any other framework. The loss functions are defined in the Supplementary Material.

*U-Net.* U-Net models are composed of an encoder and a decoder. The encoder extracts features of different scales of the RGB image through continuous down-sampling operations. The up-sampling operation of the decoder restores the feature map to the original image size with more spectral bands. U-Net generally takes RGB images as input and maps them to high-dimensional spectral images through complex transformations within the encoder and decoder. However, most methods merely focus on the spatial information, and the spectral features are usually ignored or treated as another spatial dimension; hence, the spectral reconstruction accuracy is not guaranteed.

SRUNet.   Most prior-based methods establish the RGB-to-hyperspectral mapping pixel by pixel, such as Sparse Coding and SR manifold mapping methods. These methods only consider the prior knowledge in the spectral domain and ignore the spatial context information. The authors use modified U-Net as the main framework to create SRUNet[55]. The overall network architecture diagram is shown in Fig. 3. SRUNet removes all pooling layers to avoid the loss of feature information to ensure the reconstruction results. The model focuses on local context information to enhance the spatial reconstruction results. During the training, the network accepts 32 × 32 image patches as input to further enforce this, and takes the color error metric as the objective function $L_{\Delta E}$ to reduce the spectral errors.

SRMSCNN.   To solve the SR problem, local and non-local similarity of RGB images can be useful to improve the spatial reconstruction accuracy. The authors propose a multi-scale CNN based on U-Net called SRMSCNN[35], as seen in Fig. 3. The encoder and decoder of SRMSCNN are symmetrical and connected at the same layer by skip connection operation. In the encoder part, each downsampling step consists of convolution block with max-pooling. In the decoder part, the upsampling step consists of pixel shuffle[72] (eliminating checkerboard artifacts) with a convolution block. During the training, the network takes 64 × 64 image patch as input and $\mathcal{L}_{MSE}$ as the loss function. However, this method will inevitably lose pixel information when performing downsampling by max-pooling, thereby affecting the reconstruction performance.

SRMXRUNet.   Based on the basic U-Net, the authors use the XResnet family model[73] with Mish activation function[74] as an encoder. The decoder is based on the structure proposed by Howard and Gugger's[75], while the difference is made in sub-pixel up-sampling convolution, a blur layer, and a self-attention layer. Therefore, this model is called SRMXRUNet[56], as shown in Fig. 3. The modification on the decoder reduces the loss of pixels so that more information can be kept. The added self-attention layer[76] helps the network pay attention to the relevant parts of the image. Furthermore, this network uses the improved perceptual loss function $\mathcal{L}_{mper}$[77]. All these improvements enhance the learning ability of the network, thus improving the reconstruction accuracy.

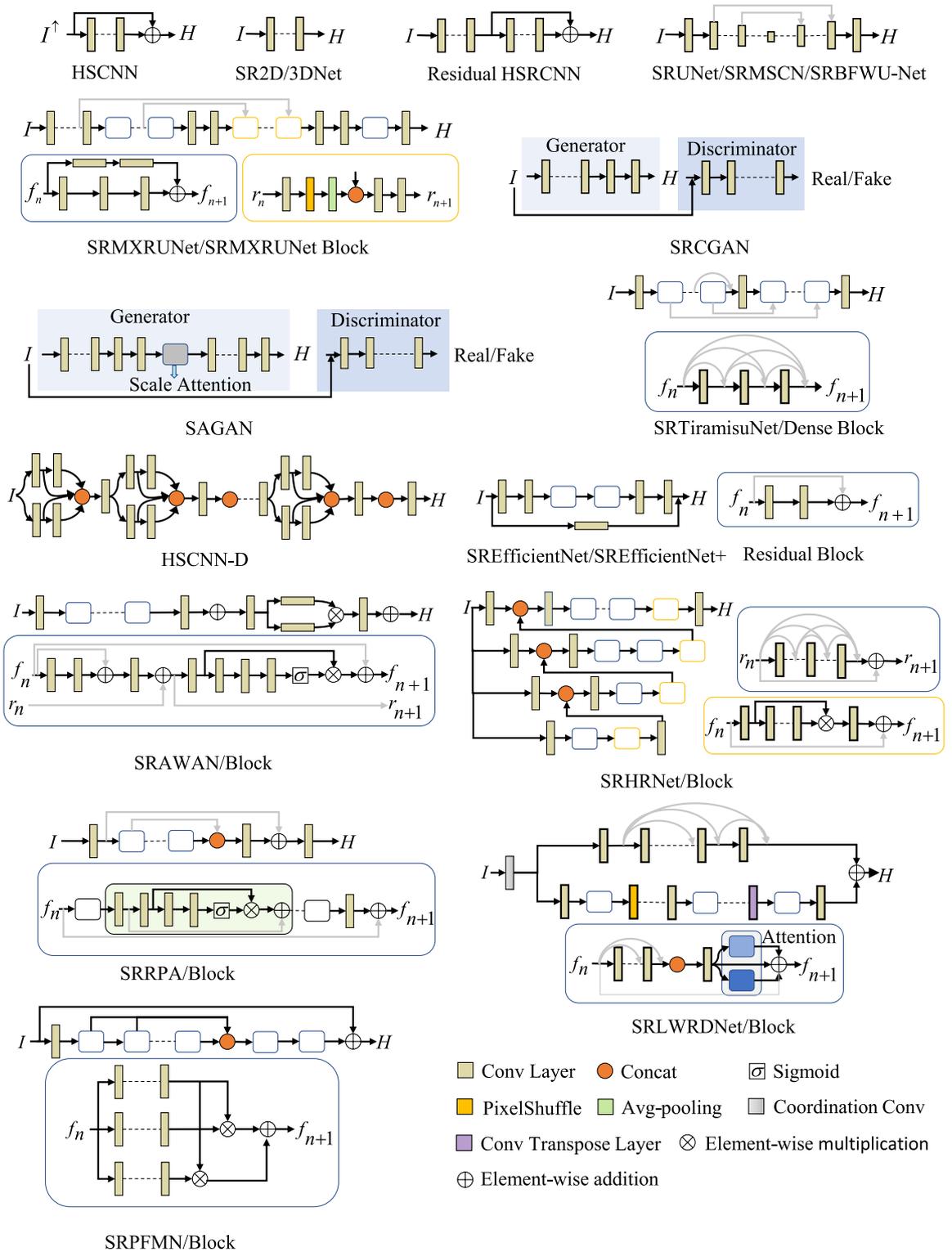

**Figure 3.** A glimpse of various deep neural network architectures used for spectral reconstruction from RGB images. $I^{\uparrow}$ represents the RGB image is upsampled in the spectral domain so that the 3D dimension is consistent with the $H$.

By adopting modified network architectures and modified loss functions, it also increases the complexity of the network, as shown in Table 3. If trained in a small dataset, the network is prone to be overfitting.

**SRBFWU‑Net.** In reality, any spectrum is a weighted combination of a set of basis spectra or functions. Some researchers found that 10 basis functions are needed to accurately generate fruitful spectral features[78–81]. Accordingly, SRBFWU‑Net[57] learns a set of 10 basis functions and related weights pixel by pixel to form a complete HSI datacube. In this way, the basis functions and corresponding weights can be continuously improved via learning, avoiding directly solving a 3-to-many ill-posed mapping problem. Besides, instead of pure supervised learning, this work introduced unsupervised learning into SR methods, which had been rarely seen in any report. In this network, two extra modules have been added to enable unsupervised learning, which are the image generation module and the photometric reconstruction loss module, as shown in Fig. 3. However, this unsupervised learning only works for RGB images obtained with known SRFs, and its effectiveness on arbitrary images using estimated SRFs might be degraded.

*GAN.* The generative adversarial network (GAN) model[82] is composed of a generator and a discriminator. The generator acts like a CNN to form a reconstructed HSI, and the discriminator judges if the reconstructed HSI is distinguishable from the ground truth or not (real or fake). These two parts battle with each other until the reconstructed HSI cannot be discriminated from the ground truth. In this manner, the network can recover hyperspectral images with better quality.

SRCGAN. The SRCGAN[32] uses conditional GAN[83] to enhance spatial context information. This network follows a basic GAN structure with a generator and a discriminator, as shown in Fig. 3. The generator is based on U-Net, and the batch normalization layer[84] is removed. PatchGAN[85] is used as the discriminator, which consists of five consecutive 3 × 3 convolutional layers. Two pairs of images $[I, H_{SR}]$ and $[I, H]$ are inputs of the discriminator. SRCGAN combines GAN loss $\mathcal{L}_{GAN}$ and $\mathcal{L}_1$ as the loss function to preserve the global structure of the image and reduce artifacts or blur[86,87]. Through the GAN structure, this method can learn the spatial-spectral distribution among channels, thereby generates high-quality spectral images. However, the network training is not always stable.

SAGAN. Two improvements have been proposed in this GAN model. The scale attention pyramid U-Net (SAP-UNet)[58] uses a U-Net with dilated convolution as the generator, and its encoding stage consists of five large residual blocks, as shown in Fig. 3. The SAP-UNet builds the feature pyramid from feature maps of the last three scale blocks, and uses the scale attention module for scale feature selection[88,89]. SAP-WNet establishes a boundary supervision branch on the basis of SAP-UNet. The improved PatchGAN is used as the discriminator of SAGAN. In this SAGAN, the loss function $\mathcal{L}_{adv}$ is adopted from the objective function of WGAN[90], which is further combined with the $\mathcal{L}_1$ to form the total loss function. This modification on loss function can overcome the training instability issue. Moreover, adding dilated convolution and attention mechanism to enrich the network structure improves the network performance. However, the downsampling in the generator might lead to loss of pixel information.

*Dense network.* The core idea of dense network[91] is to densely connect all layers to achieve higher flexibility and richer feature propagation, which can alleviate the vanishing-gradient problem and promote the stability of the network.

SRTiramisuNet. The SRTiramisuNet[59] adopts a variant of the Tiramisu network[92], which is a type of dense network, as shown in Fig. 3. More importantly, its architecture is based on a multi-scale paradigm, allowing the network to learn the overall image structure while keeping the image resolution fixed. This SRTiramisuNet has an input of 64 × 64 image patch, and takes Euclidean loss $\mathcal{L}_{eu}$ as the objective function. As a dense network, this method has advantages of less parameters, high efficiency in training and robustness. However, the downsampling by max-pooling cannot ensure a high reconstruction accuracy.

HSCNN+. The HSCNN+[60] has three variants based on HSCNN, namely HSCNN-U, HSCNN-R, and HSCNN-D. HSCNN-U uses 1 × 1 convolutional layer to achieve spectral upsampling, which has slightly improved the performance. HSCNN-R replaces the plain convolutional layer of HSCNN with residual blocks while remaining global residual learning to further improve the accuracy. HSCNN-D replaces residual blocks by dense blocks with the path-widening fusion scheme, which can substantially alleviate the vanishing of gradients issue. The model uses the $\mathcal{L}_{MRAE}$ loss function, and takes 50 × 50 RGB image patches as input. Compared to HSCNN-R, HSCNN-D takes feature fusion as a concatenation block instead of an addition (Fig. 3) to achieve a deeper network. Such that, it can learn a better mapping relation, and provide higher reconstruction fidelity. On top of HSCNN, this work eliminates the strong dependence on the SRF. Nevertheless, it is difficult to achieve a balance between speed and performance.

*Residual network.* Residual network adds skip connections among layers and adopts deeper blocks to avoid the vanishing of the gradient. The residual network not only makes the reconstructed image more detailed, but also maintains low-frequency information well.

SREfficientNet. The SREfficientNet[61] employs residual blocks to make full use of low-level features. The network mainly contains a backbone network and a convolutional layer, as shown in Fig. 3. The convolutional layer learns the residual mapping from RGB to hyperspectral. The backbone network sequentially performs like three modules, shallow feature extraction and shrinking, complex feature extraction as well as expansion and reconstruction. The first module reduces over-fitting and forces the network to learn more compact and relevant features. In the second module, two residual blocks[93] are stacked to obtain more complex features. The third module expands the features and reconstructs the HSI. The input of the model is 36 × 36 image patches, while the network is trained with $\mathcal{L}_2$. This method adopts residual learning to avoid the vanishing of the gradient. The combined use of local residuals and global residuals expands the receptive field to enhance the feature expression ability of the network. Compared with other recent state-of-the-art methods based on CNN, this method requires much less memory, computing resources, and running time.

SREfficientNet+.    The network structure of SREfficientNet+[62] is in consistent with SREfficientNet, except that this method has a foregoing estimation block that provides a set of estimated SRF as an extra input to ensure the consistency of the spectral reconstruction process. As a result, SRefficientNet+[62] further advances towards the SR problem for RGB images acquired under unknown cameras (or estimated SRFs), enabling the network training using more versatile datasets.

*Attention network.*    The previously discussed networks usually treat spatial and spectral information equally. In some cases, we have to selectively focus on certain features in a given layer. Attention networks allow this flexibility and consider that not all features are important for SR by adding the attention mechanism.

SRAWAN.    The adaptive weighted attention network for SR (SRAWAN)[63] explores the camera spectral sensitivity (CSS) prior and the interdependence among intermediate features to promote the reconstruction accuracy. SRAWAN is composed of a convolutional layer, dual residual attention blocks (DRAB)[94] and patch-level second-order non-local module (PSNL)[95]. The convolutional layer performs the functions of shallow features extraction, deep feature extraction and reconstruction, as shown in Fig. 3. Besides, stacking several DRAB benefits deeper features extraction. Here this attention module is inherited from the Squeeze-and-Excitation[96], while the difference is that the adaptive weighted feature statistics (convolutional layer) replaces the global average pooling statistics, to strengthen feature learning. PSNL can capture long-range spatial context information to further improve the reconstruction of spectral images. The loss function is a combination of the CSS loss $\mathcal{L}_{CSS}$ and $\mathcal{L}_{MRAE}$. This method integrates various strategies, such as attention mechanism and residual learning, which greatly increases the feature representation ability, thereby obtaining high-accuracy spectral images. Given that this method still needs camera sensitivity as a prior to improve accuracy, its application in more practical cases is limited.

SRHRNet.    The SRHRNet[64] is a 4-level hierarchical regression network, where each level consists of residual dense blocks[93,97] and residual global blocks[93,96], as seen in Fig. 3. The residual dense blocks can greatly degrade the artifact effect. The residual global blocks have skip connection from the inputs to extract attentions or features for every long-range pixels through attention mechanism. Both PixelUnShuffle and PixelShuffle are employed in the downsampling and upsampling steps respectively, which reduces noise in the generated image and reserves detailed information. The integration of multiple modules at the top of the network can effectively obtain high-quality reconstructed HSIs. SRHRNet uses $\mathcal{L}_1$ as the loss function and proposes an 8-setting ensemble strategy to further enhance generalization.

SRRPAN.    The residual pixel attention network (SRRPAN)[65] is proposed to adaptively rescale each pixel-wise weights of all input feature maps, as shown in Fig. 3. SRRPAN contains a residual attention group (RAG) and a residual attention module (RPAB). Each RAG is composed of RPAN blocks and a convolutional layer, all with residual learning. RPAB contains pixel attention (PA) block with skip connection from the input to obtain pixel-level attention features. As the network depth increases, each RAG extracts features of different scales[98]. In order to make full use of these feature maps, they are fused through the concat layer to improve the quality of the reconstructed HSI. During the training, the network takes 64 × 64 RGB image patches as inputs and $\mathcal{L}_{MRAE}$ as the loss function. This method proposes the pixel attention mechanism, so that the network pays more attention to the features of different locations. However, the context information is ignored, thus cannot ensure a high reconstruction accuracy.

*Multi-branch network.*    In contrast to single-stream CNN, the goal of multi-branch network is to obtain a diverse set of features on multiple context scales. This information is then fused to obtain better reconstructed HSI. This design can also achieve multi-path signal flow, leading to better information exchange during training. We explain several SR multi-branch networks below.

SRLWRDANet.    The SRLWRDANet[66] consists of two parallel subnets (branches), which are a densely connected network and a multi-scale network, as illustrated in Fig. 3. The network starts with a coordinated convolution block[99] to extract shallow features and boundary information, the parameters of which are shared by the latter two subnets. The densely connected structure of the network enables a strong feature representation ability and effectively alleviates the vanishing of gradient. The multi-scale network is composed of a convolutional layer and several residual dense attention blocks (RDAB)[100] to extract various scale-level features. The subnet has a multi-scale connection of RDAB in the U-Net fashion, where the downsampling is implemented by the max-pooling and the deconvolution completes the up-sampling. SRLWRDANet takes the sum of $\mathcal{L}_2$ and structural similarity loss ($\mathcal{L}_{SSIM}$) as the objective function to achieve the purpose of preserving structural features. Notice that the usage of branch structure is effective to extract features of different scales, which should be quite helpful for promoting learnable SR methods.

| Category | Method | BGU-HS | | | ARAD-HS | | |
|---|---|---|---|---|---|---|---|
| | | RMSE | MRAE | SAM | RMSE | MRAE | SAM |
| Dictionary learning | Sparse coding | 51.48 | 0.0808 | 5.01 | 0.0331 | 0.0787 | 6.46 |
| | SR A+ | 26.09 | 0.0448 | 2.83 | 0.0226 | 0.0725 | 4.61 |
| Linear CNN | HSCNN | 17.006 | 0.0190 | – | – | – | – |
| | SR-2DNet | 21.394 | 0.020 | – | – | – | – |
| | SR-3Dnet | 20.010 | 0.018 | – | – | – | – |
| U-Net | SRUNet | 15.88 | 0.0156 | 1.11 | 0.0152 | 0.0395 | 2.74 |
| | SRMSCNN | 19.28 | 0.0231 | 1.47 | 0.0235 | 0.0724 | 4.91 |
| | SRMXRUNet | – | – | – | – | 0.0454 | – |
| | SRBFWU-Net | – | – | – | 0.0151 | 0.0434 | – |
| Dense network | SRTiramisuNet | 20.98 | 0.0272 | 1.57 | 0.0251 | 0.0850 | 4.34 |
| | HSCNN-R | 13.911 | 0.0145 | 1.05 | 0.0143 | 0.0372 | 2.63 |
| | HSCNN-D | 13.128 | 0.0135 | **0.99** | – | – | – |
| Attention network | SRAWAN | **10.24** | **0.0114** | – | **0.0111** | **0.0312** | **2.16** |
| | SRHRNet | – | – | 1.01 | 0.0135 | 0.0423 | 2.53 |

**Table 4.** Reconstruction accuracy comparison of representative SR methods in terms of RMSE, MRAE and SAM on the BGU-HS and ARAD-HS datasets. Top two best results are highlighted in bold and underline respectively.

**SRPFMNet.** Most SR methods solves the RGB-to-hyperspectral mapping in a size-specific receptive field centered on a certain pixel. Because of their different category and spatial position, pixels in hyperspectral usually require different sized receptive fields and distinct mapping functions. The Pixel-aware deep function-mixture network (SRPFMNet)[67] based on multi-branch network is proposed to solve this problem. The SRPFMNet is composed of a convolution layer followed by Relu and multiple function mixing (FM) modules, as seen in Fig. 3. It fuses the intermediate features generated by the previous FM blocks with skip connection, while adopting global residual structure. Each FM module includes mixing function and several basis function subnets, and these networks are formed by stacking multiple convolutional layers. The mixing function subnet generates pixel-wise mix weights. These basis function subnets have different-sized convolution kernels to generate receptive fields of different sizes and learn distinct mapping schemes. The models use $\mathscr{L}_1$ for the training and a patch size of 64 × 64 as the input. From this method, the attention strategy to different positions and different types of pixels can improve the reconstruction accuracy and quality.

## Comparison and analysis

We compare the spatial and spectral performance of different SR algorithms based on the BGU-HS[27] and ARAD-HS[28] datasets. The results are summarized in Table 4. Regarding the datasets, we have kept the original resolutions and the image amounts. As BGU-HS dataset has a smaller amount and much higher resolution, it shows a much worse convergence compared to the ARAD-HS dataset. Furthermore, statistics show that (1) in general, the dictionary methods show much larger reconstruction errors compared to most deep-learning methods; (2) the reconstruction performance of the SRAWAN algorithm is the best among all the trained methods in terms of RMSE, MARE and SAM. Both RMSE and MARE are criteria from the perspective of spatial reconstruction errors, while the SAM reflects the spectral reconstruction errors between the reconstructed and the original spectral images. As most of above-mentioned methods have originated from computer vision algorithms, the spectral accuracy is considerably overlooked. Table 4 clearly shows that data-driven methods can obtain better spectral quality than prior-based methods. However, those earlier deep learning-based methods (like SRMSCNN and SRTiramisuNet) pay more attention to the geometric properties of the HSIs, which results in good reconstructed 2D image but poor spectral properties. In addition, recent methods (e.g., SRHRNet) learn also the space-spectral relations to improve the quality of the reconstructed spectra.

Figures 4 and 5 are used to visualize the reconstruction performance from the spatial 2D image and the 1D spectral curve domains respectively. The HSIs of the simulation datasets all have 31 channels, and we notice all of them have similar errors. In Fig. 4, we pick the 640 nm-channel to present the reconstruction errors of selective methods in heat maps, resembling the MRAE of the reconstructed 2D image. It is clear that the sparse coding method has much larger errors across the whole image, while the SRAWAN method shows negligible errors. Figure 5 shows the spectral curves of two spatial points on the reconstructed HSIs and the original HSIs. The spectral curve using sparse coding has a much larger difference from the ground truth curve than the other methods. The data-driven method with attention mechanism and residual learning (SRAWAN) achieves the smallest difference from the ground truth spectral curve on both points.

**Prior-based methods.** In general, the prior-based methods combine known prior knowledge to improve the reconstruction performance. More specifically, the sparse coding method considers only the sparsity of HSIs, while other methods in dictionary learning incorporate more priors such as spectral feature correlation, spatial context information, and local linear relationship to improve the representation ability of the dictionary. On top of the known priors, manifold learning and Gaussian process build models based on the statistical features (e.g., low-dimensional manifold and non-negativity of spectra) of HSIs.

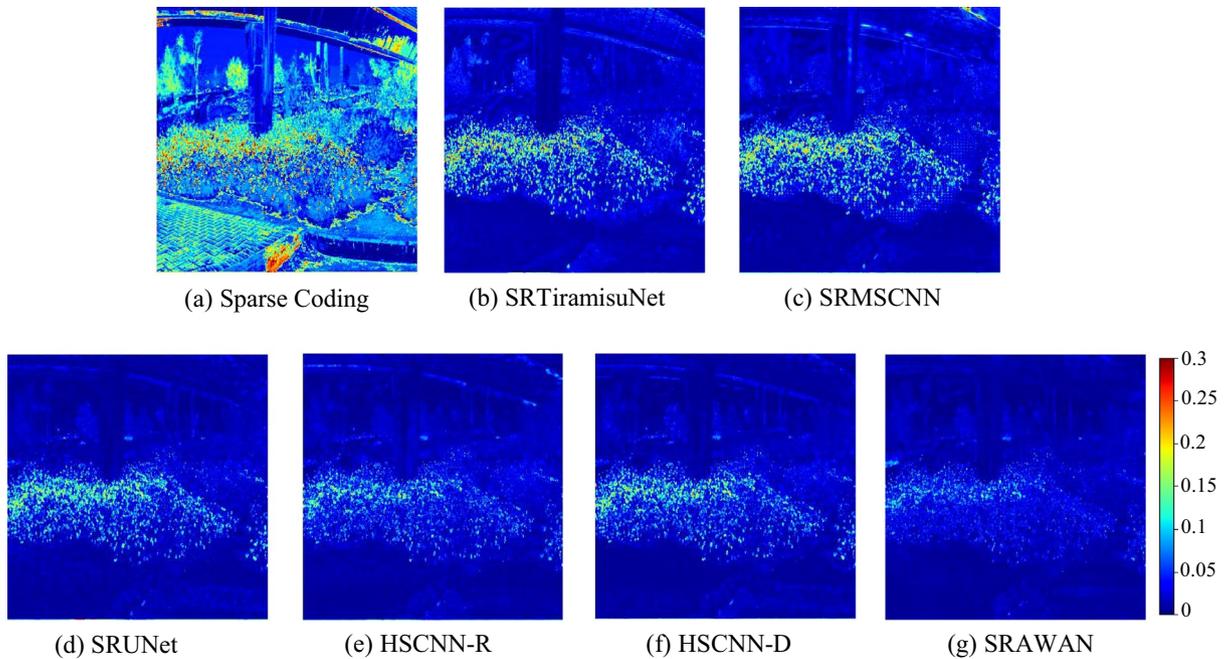

**Figure 4.** Performance comparison of selective SR methods using the residual heat map at 640 nm channel. The ground truth image is from the BGU-HS open-source dataset[27]. MRAE errors of Sparse Coding[29] (**a**), SRTiramisuNet[59] (**b**), SRMSCNN[35] (**c**), SRUNet[55] (**d**), HSCNN-R[60] (**e**), HSCNN-D[60] (**f**), and SRAWAN[63] (**g**) in the spatial domain.

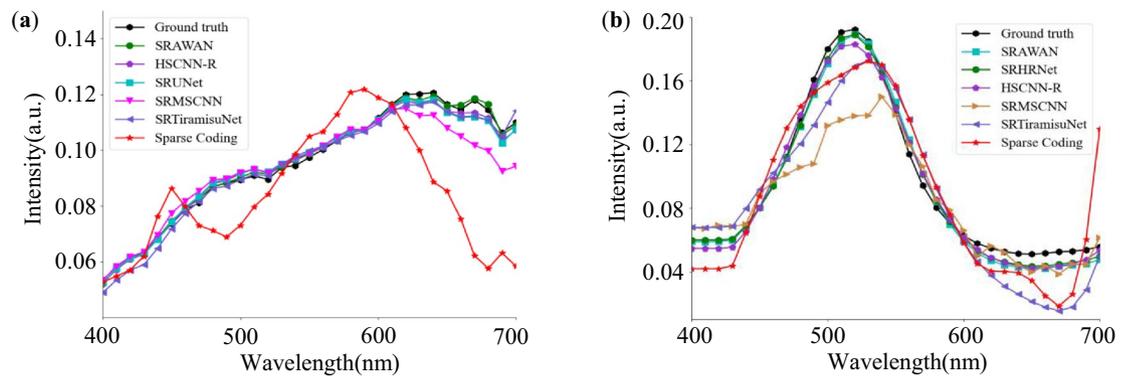

**Figure 5.** Spectral error comparison of SR methods on selected spatial locations. Data from BGU-HS (**a**) and ARAD-HS (**b**).

Nevertheless, these prior-based methods ignore the characteristics of spatial structure similarity and correlation between spectra, leading to weak representation and loss of high-frequency information. Besides, the performance of most such algorithms depends on handcrafted priors, making the representation ability of their dictionaries vulnerable to some underestimated hypotheses. In most circumstances, the SRF should be known to establish a dictionary for learning the RGB-to-spectral mapping. In the case of estimated SRFs, the reconstruction performance and accuracy will be degraded, indicating poor portability and compatibility.

**Data-driven methods.** Data-driven methods leverage the large amount of available RGB images to find the hidden mapping relations, thus predict more accurate HSIs. The natural images have rich spectral structure information that are not obvious enough to generate known priors, while these can be well matched with the powerful feature representation ability of neural networks.

So far, there are many SR methods based on CNN. Linear CNN is the earliest SR model with a simplified network architecture and a moderate performance, as shown in Table 4. The GAN models are one of the complex structures based on CNN, e.g., SRCGAN, SAGAN. Through sufficient iterative adversarial training, the generator can produce output consistent with the ground truth HSIs, while the discriminator cannot distinguish between the reconstructed HSIs and the ground truth HSIs. The U-Net based model makes use of downsampling, so the receptive field of the network is constantly increased, which makes the network perceive more pixels e.g., SRUNet, SRMSCNN, SRBFWU-Net, SRMXRU-Net. By doing so, the local and non-local information of the image can be jointly encoded, which can slightly improve performance, as shown in Table 4. The residual networks and dense networks can alleviate the vanishing of gradients during the training, thus more stable and higher accurate results can be obtained. The attention networks and multi-branch networks greatly enhance SR performance by increasing network complexity. Statistics from Table 4 show that the algorithms with attention mechanism (SRAWAN and SRHRNet) have the best performance among others. In this sense, more advanced attention blocks can be further explored in the future to enhance the learning ability and reconstruction accuracy, such as spatial and channel mixed attention[101] and layer attention[102], etc.

Note that, it is difficult to make a fair comparison and find a winner among those reconstruction methods based on deep learning. Apart from the metrics of reconstruction accuracy and generalization, there are many other influential factors, such as time cost, consistency and network complexity.

**Challenges and trends.** In previous sections, we describe many state-of-the-art methods with simulation results on their performance both from spatial and spectral reconstruction accuracy. In short, the prior-based methods are limited by a few known hand-crafted features (priors), thus tend to have worse reconstruction performance. Moreover, with the rapid increment of hyperspectral datasets, the data-driven methods can perform better by continuous training, thus becomes more promising. Although current spectral reconstruction methods based on deep learning have been making huge progress, there are still many challenges for future development of such algorithms.

*Network architecture.* Spatial context information plays an important role in the network performance, so both local and global information could be considered in building the network architecture. Low-frequency and high-frequency information determine the quality of the reconstructed image, and this information should also be included into the network design. In different scenarios, people often pay attention to different features of things, and combining attention mechanisms of key features can enhance the production of interested details. The combination of hierarchical structure and attention mechanism can make the network more enhanced with feature representation ability, e.g., SRHRNet. Besides, non-local modules can also greatly improve the learning ability of the network, e.g., SRAWAN.

*Loss function.* A suitable loss function of the neural network is critical for obtaining the optimal solution in the huge solution space. Existing network models use pixel-level loss functions, such as $\mathscr{L}_2$ and $\mathscr{L}_1$. They are typically applied to RGB image processing, but they cannot describe both the spatial and spectral features of HSIs very accurately. Therefore, various loss functions (e.g,. content loss, perceptual loss and texture loss) could be firstly exploited to improve the reconstruction accuracy in the spatial domain. More importantly, as the applications of HSIs greatly depend on the accurate spectra, the above-mentioned loss functions can lead to the metamer problem[43], meaning the same RGB pixel maps to different spectral curves. Thus, some advanced loss functions that can describe the spectral loss could be developed to solve this metamer problem from the perspective of spectral accuracy. It seems using the among-channels $\mathscr{L}_{MRAE}$ as the loss function can effectively reduce the metamer issue, for example in the SRAWAN and SRBFWU-Net methods.

*Datasets.* Nowadays, available hyperspectral images are still very limited, while deep learning training often requires a large amount of data. Thus, existing methods are most likely to be over-fitted. Given that conventional hyperspectral acquisition is relatively expensive, many image augmentation strategies can be used to increase the training datasets, such as cropping, flipping, zooming, rotating, and color dithering. Another interesting strategy is to combine prior knowledge with deep learning methods to reduce the requirements on large datasets. Pioneer work has been reported from Wang et al.[103] by constructing a joint prior-based and data-driven spectral reconstruction network. Except for the rare amount in current HSIs datasets, those input RGB images are usually synthesized from hyperspectral images by the CIE color matching function[104], which are different from RGB images captured from real scenes. Therefore, the SR networks that are trained under such assumptions might not be well applied to actual RGB images. Tentative solutions to this problem are to use unsupervised learning that directly inputs real RGB images, or to ensure the synthetic RGB images have negligible difference from real images.

*Unsupervised learning.* Almost all the methods in this paper are based on supervised learning on the condition that the ground-truth HSIs and the corresponding RGB images are known. However, such datasets are still quite rare, much more difficult and expensive to build than common RGB image datasets. Not to mention that in some circumstances the HSI datasets are still not possible, like the endoscopic images. To solve the issue of rare qualified datasets, SRBFWU-Net explored unsupervised spectral reconstruction and achieved quite encouraging results (Table 4). The development of unsupervised learning using more abundant RGB datasets might greatly prompt and broaden the applications of SR methods in other fields.

*Portability and compatibility.* Currently, most methods perform poorly on spectral reconstruction from RGB images acquired with unknown camera parameters. This might limit the applications of those methods to more practical situations. In addition, many HSIs often have hundreds of spectral bands (e.g., 256), much more than 31 bands that are used to train all the SR methods here. Under such extreme spectral upsampling conditions, it could be very challenging to maintain a high accuracy in spectral domain, and the methods might not be compatible anymore. Future approaches should consider these issues to increase the network portability and compatibility.

## Conclusion

We have performed a systematic review of spectral reconstruction algorithms from RGB to hyperspectral images, categorized into prior-based methods and data-driven deep learning approaches. The mathematical relationship between RGB images and hyperspectral images is given as the fundamentals for the reconstruction methods. We have summarized the features of four available HSI datasets and listed the metrics for evaluating the reconstruction accuracy. The amount and image resolution of the datasets show considerable impacts on the reconstruction accuracy. When the dataset is small, prior-based methods could be chosen, as deep learning methods may suffer from overfitting. However, the prior-based methods rely heavily on handcrafted priors and known SRFs. When the dataset is large, data-driven deep learning methods is preferred, as they tend to better map the RGB images to the real HSIs. Based on two typical datasets, we have trained and compared several selected deep-learning methods to give a thorough guidance on which mechanisms are better for such networks.

In the end, we summarize the challenges of current reconstruction methods to enlighten some potential research topics. One of the most crucial improvements must be the spectral reconstruction accuracy, usually ignored but essential to promote the application of such methods. Spectral loss functions could be devised to better recover the spectral features and avoid large consistency errors. Secondly, statistics on our simulation results show that advanced network architectures like the attention mechanism, multi-branch and hierarchical structures can improve the overall reconstruction accuracy and compatibility to more practical situations. Other advanced networks could also be investigated, such as Transformer[105,106]. Almost all data-driven methods are based on supervised learning, while the full potential of this technique is to be unleashed by unsupervised learning. The expansion of HSI datasets is necessary and urgent, very important to improve both the reconstruction accuracy and the algorithm robustness. With those practical concerns solved, the spectral reconstruction algorithms will eventually be applied in more and more consumer-level fields.

## Data availability

The datasets generated and/or analysed during the current study are available in the Spectral Reconstruction repository. See https://github.com/Intelligent-Imaging-Center/Spectral-Reconstruction.

### Acknowledgements
This work was supported by the Chinese Academy of Sciences (No. CAS-WX2021-PY-0110, NO. YJKYYQ20180039 and NO. Y70X25A1HY ), and the National Natural Science Foundation of China (NO. 61775219, NO. 61771369 and NO. 61640422). Yunfeng Nie acknowledges the Flemish Fund for Scientific Research (FWO) for supporting her research (No. FWOTM1039). The authors would like to thank those who provide open-sources code for the whole community, such as Sparse Coding, SRUNet, SRMSCNN, HSCNN+ and etc.

### Author contributions
Y.N., Q.F. and J.Z. proposed the concept. R.S. and Q.F. investigated the background. R.S. performed the simulations and collected the results. Q.F., W.R. and F.H. guided the simulations and checked the results. J.Z. and Y.N. supervised the entire work. All authors wrote and revised the manuscript.

### Competing interests
The authors declare no competing interests.

### Additional information
**Supplementary Information** The online version contains supplementary material available at https://doi.org/10.1038/s41598-022-16223-1.

**Correspondence** and requests for materials should be addressed to Y.N.

**Reprints and permissions information** is available at www.nature.com/reprints.

**Publisher's note** Springer Nature remains neutral with regard to jurisdictional claims in published maps and institutional affiliations.